\documentstyle[prl,aps,preprint,epsf]{revtex}

\begin{document}
\input psbox.tex
\newcommand{\be}{\begin{equation}}
\newcommand{\ee}{\end{equation}}
\newcommand{\bq}{\begin{eqnarray}}
\newcommand{\eq}{\end{eqnarray}}
\newcommand{\Sc}{Schr\"odinger\,\,}
\newcommand{\Sp}{\,\,\,\,\,\,\,\,\,\,\,\,\,}
\newcommand{\no}{\nonumber\\}
\newcommand{\tr}{\text{tr}}
\newcommand{\p}{\partial}
\newcommand{\la}{\lambda}
\newcommand{\G}{{\cal G}}
\newcommand{\D}{{\cal D}}
\newcommand{\W}{{\bf W}}
\newcommand{\de}{\delta}
\newcommand{\al}{\alpha}
\newcommand{\bi}{\beta}
\newcommand{\ga}{\gamma}
\newcommand{\ep}{\epsilon}
\newcommand{\vep}{\varepsilon}
\newcommand{\th}{\theta}

\setcounter{page}{0}
\def\footnoterule{\kern-3pt \hrule width\hsize \kern3pt}
\tighten
\title{
Static Colored $SU(2)$ Sources in $(1+1)$-Dimensions - An Analytic Solution
in the Electric Representation
\footnote{This work is supported in part by funds provided by the 
U.S. Department of Energy (D.O.E.) under cooperative research agreement 
\#DF-FC02-94ER40818.
}
}
\author{Jiannis Pachos, 
\footnote{Email address: {\tt pachos@ctp.mit.edu}}
Antonios Tsapalis
\footnote{Email address: {\tt tsapalis@ctp.mit.edu}}}
\address{Center for Theoretical Physics \\
Massachusetts Institute of Technology \\
Cambridge, Massachusetts 02139 \\
{~}}

\date{MIT-CTP-2681, October 1997}
\maketitle

\thispagestyle{empty}

\begin{abstract}

Within the \Sc Electric Representation we analytically calculate 
the complete set of wave functionals obeying Gauss' law 
with static sources in a general representation of $SU(2)$ 
in $(1\!+\!1)$-dimensions.
The effective potential is found to be linear in the distance between 
the sources with the string tension depending on the polarization of the 
solution.
\end{abstract}

\vspace*{\fill}

\pacs{xxxxxx}

\section{Introduction}

In $(1\!+\!1)$-dimensional Yang-Mills theory the Hamiltonian is non-dynamical since there are no transverse degrees of freedom. For the temporal gauge the wave functional is determined from the Gauss' law constraint. In the case that we insert into the theory static sources, the Hamiltonian remains the same, but the Gauss' law is appropriately modified. The Wilson line of the vector potential is the only known solution satisfying the modified Gauss' law in the coordinate Schr\"odinger representation. In this work we deal with the problem in the momentum representation and we derive all the solutions for static sources in a general representations of $SU(2)$. In particular, for sources in the $j$ representation, $2j+1$ independent solutions are found and the corresponding string tension is calculated. It is not possible to arrive at these solutions by straightforwardly Fourier transforming the Wilson line solution of the coordinate representation. However, we obtain that a superposition of our solutions possesses the same energy eigenvalue as the Wilson line.

\section{Schr\"odinger Representation}

The action of pure Yang-Mills is given by
\be
S:=-{1 \over 2} \int dx dt \,\, \tr \left( F_{\mu \nu} F^{\mu \nu}\right)\,\,,
\ee
where $\mu$, $\nu$ run from $0$ to $1$ and $F$ is the field strength given by
\be
F_{\mu \nu}=\p_\mu A_\nu-\p_\nu A_\mu-i[A_\mu,A_\nu]\,\, ,
\ee
where $A_\mu=A^a_\mu T^a$, 
with hermitian matrices, $T^a$, belonging 
to a Lie Algebra and satisfying 
$[T^a,T^b]=if_{abc}T^c$ 
and $\tr (T^a T^b)=c \de^{ab}$, for $c$ a positive number.
We have set the coupling constant $g$ equal to one.
Here we shall take $T$ to belong in some representation of $SU(2)$. 
As we want to work in the Hamiltonian formalism, we choose the 
temporal gauge, $A_0=0$, so that no time dependent gauge 
transformations are allowed.
We can set $A_1^a \equiv A^a$ and $E^{1a} \equiv E^a$.
The Hamiltonian is
\be
H={1 \over 2}\int dx (\p_0 A(x))^2 \,\,.
\ee
We can choose the \Sc representation to materialize the 
quantization condition
\be
[\hat A^a(x),\hat E^b(y)]=i\delta^{ab}\delta(x-y)
\label{qua}
\ee
for $E(x)$ being classically the ``electric'' field of the theory, i.e. 
\be
E^{1a}(x)=F_{01}^a(x)=\p_0A_1^a(x)-\p_1 A_0 ^a(x)+f_{abc}A_0 ^b(x)A_1 ^c(x)
\ee
which in the temporal gauge becomes $E^a(x)=\p_0A^a_1(x)$.
The quantization condition (\ref{qua}) can be fulfilled for 
$\hat E^a(x)=-i{\delta \over {\delta A^a(x)}}$ and 
$\hat A^a(x)=A^a(x)$, that is the ``A''-representation, or by
$\hat A^a(x)=i{\delta \over {\delta E^a(x)}}$ and
$\hat E^a(x)=E^a(x)$ which is the ``E''-representation.
The Gauss operator in the temporal gauge is given by
\be
\hat \G_a(x)={\p \over {\p x}}{\delta \over {\delta A^a(x)}} +
f_{abc}A^b {\delta \over {\delta A^c(x)}} \,\, .
\ee
This operator acts on the functional space of states, an element 
of which can be written as
$\Psi[A]$. The Gauss' law which results as a constraint from the 
equations of motion of the theory reads
\be
\hat\G_a \Psi[A]=0 \,\, .
\label{ga}
\ee
In the case we want to insert n static point-like sources in the theory
the Gauss' law is modified to the following
\be
\hat \G_a \Psi[A]=-i\sum_{\al=0}^{n-1}T^{(\al)}_{\,a} 
\delta(x-x_\al)\Psi[A]
\label{gap}
\ee
where $x_a$ are the spatial coordinates of the sources.
The functional Fourier transformation of a general functional is defined as 
\be
\tilde \Psi[E]= 
\int {\cal D}A \, \exp\left( - i \int E_a A_a\right)\Psi[A]
\ee
Using the inverse of this definition we can find the action of the Gauss 
operator on a functional, $\tilde \Psi[E]$, to be
\be
\left( {\p \over \p x} E^a -i f_{abc} E^b 
{\delta \over \de E^c}\right) \tilde \Psi[E]=
-\sum_{\al=0}^{n-1} T^{(\al)}_{\,a}  \delta(x-x_\al)\tilde \Psi[E]\,\, .
\label{gape}
\ee
The Hamiltonian operator in the ``A''-representation is
\be 
\hat H= {1 \over 2} \int dx \left(-i{\de \over {\de A^a}} \right)^2
\label{arep}
\ee
and the ``E''-representation is given by
\be
\hat H= {1 \over 2} \int dx \left(E^a \right)^2
\ee
that is it acts multiplicatively. We see that the Hamiltonian equation
does not describe {\it any} dynamics, so the wave functional 
needs to satisfy only the Gauss' law.

For example for the free $SU(N)$ case the Gauss' law without sources is
\be
\hat \G_a \Psi[E]=
\left( {\p \over \p x} E^a -i f_{abc} E^b 
{\delta \over \de E^c}\right) \Psi[E]=0
\label{zag}
\ee
and the wave functional satisfying it is \cite{Jack1} \cite{Jack2}
\be
\Psi [E] =\prod _x \delta(|E(x)|-\rho) \exp \left(
-{1 \over c}\int dx \,\, \tr(E g^{\prime} g^{-1}) \right)
\ee
where $|E|$ is the magnitude of the electric field, $\rho$ is a constant
number, g is an element of the gauge group $SU(N)$ defined from the 
general decomposition of $E$ as $E=gKg^{-1}$, 
$K$ belongs to the corresponding algebra, and the
prime stands for differentiation with respect to space. 
As $g^{\prime}g^{-1}$ is always anti-hermitian the exponential
is a phase factor.
What the delta function does is to make the magnitude of the 
electric field constant with respect to $x$.
This can be seen when (\ref{zag}) is
contracted with $E^a$ and the antisymmetry of $f_{abc}$ is used. 
This is equivalent to having the matrix $K$ independent of $x$. Also
$\tr K^2=\tr E^2= c \rho ^2$.

\section{Wave Functional}

\begin{figure}
\centerline{
\put(17,60){$1$}
\put(90,20){$x_0$}
\put(220,20){$x$}
\psannotate{\psboxscaled{500}{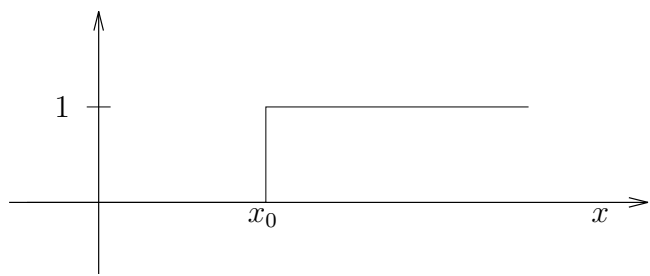}}
{\at(-9\pscm;-1\pscm)
{\caption{\label{step}Step function $\th(x-x_0)$.}}}
}
\end{figure}

We can construct the wave functional for sources, $T_a$, 
belonging in the fundamental representation of $SU(2)$.
For this purpose we use instead of the coordinates of the electric field
$E_1$, $E_2$ and $E_3$, their spherical decomposition,
\be
\begin{array}{ccc}
E_1=|E|\cos \alpha \sin \beta\Sp \Sp & & 
\Sp \Sp  E_+=E_1+iE_2=|E|e^{i\alpha}\sin \beta\\
E_2=|E| \sin \alpha \sin \beta\Sp \Sp &\text{or}&
\Sp \Sp E_-=E_1-iE_2=|E|e^{-i\alpha} \sin \beta\\
E_3=|E|\cos \beta \Sp \Sp& & 
\Sp \Sp E_3=|E|\cos \beta
\end{array}
\ee
The differential generators of the angular momentum of a point particle are
\be
J_a=-i \varepsilon_{abc} E_b(x) {\de \over \de E_c(x)}
\ee
or in the new coordinates
\bq
&&
J_+=J_1 +i J_2= e^{i \al} \left( {\de \over \de \bi} + i \cot \bi 
{\de \over \de \al} \right)
\nonumber\\
&&
J_-=J_1-i J_2= e^{-i \al} \left( -{\de \over \de \bi} + i \cot \bi 
{\de \over \de \al} \right)
\nonumber\\
&&
J_3=-i {\de \over \de \al}
\eq
In the fundamental representation we choose the $T_a$ matrices to be 
written in terms of the Pauli matrices
\be
\sigma_3=
\left(\begin{array}{cc}
1&0\\
0&-1
\end{array}
\right)\,\, ,\,\,\,\,
\sigma_+=\sigma_1+i\sigma_2=2
\left(\begin{array}{cc}
0&1\\
0&0
\end{array}
\right)\,\, ,\,\,\,\,
\sigma_-=\sigma_1-i\sigma_2=2
\left(\begin{array}{cc}
0&0\\
1&0
\end{array}
\right)
\ee
as $T_a=-i \sigma_a/2$. From the solution of the homogeneous Gauss' 
law we can guess that the wave functional satisfying
(\ref{gape}) can be of the form
\be
\Psi[E]=\prod_x \de \Big(|E(x)|-k\th(x-x_0) -\rho \Big)
\times e^{i\Omega} \times {\bf W}
\label{ans}
\ee
where $k$ and $\rho$ are constants, ${\bf W}$ is 
a $(2\! \times \! 1) $ column with entries functions of $E(x_0)$, the 
electric field at the point $x_0$, and $\th(x-x_0)$ is the step function
at the point $x_0$ given in Figure (\ref{step}).
For $\Omega$ we can take an explicit form of ${i \over c}\int dx 
\,\, \tr(E g^{\prime} g^{-1})$ as
$\Omega = \int dx |E|\cos \bi \, \al ^{\prime}$ for the matrix, $K$,
being fixed on the third direction. Substituting (\ref{ans}) into 
Gauss' law (\ref{gape}) for $f_{abc}=\vep_{abc}$ we get 
\bq
&&
\prod_x\de \Big(|E|-k \th (x-x_0)-\rho\Big) \times 
{\bf W} \times\! \left( E_a^{\prime} e^{i \Omega}
-i\vep_{abc} E_b {\de e^{i\Omega} \over \de E_c}\right)- \Sp \Sp
\no
&&
\Sp  \Sp \prod_x\de \Big(|E|-k\th(x-x_0)-\rho\Big) e^{i\Omega} 
\de (x-x_0) i\vep_{abc} E_b(x_0)
{\p {\bf W} \over \p E_c(x_0)}=
\no
&&
\Sp=-\prod_x\de \Big(|E|-k \th(x-x_0)-\rho \Big) e^{i\Omega} 
{\sigma_a \over 2 } {\bf W} \de (x-x_0)
\label{interm}
\eq
We can evaluate the first part of expression (\ref{interm}), which gives
the following components
\be
(E_3^{\prime}+J_3)e^{i\Omega}=0 \,\, , \,\,\,\, 
(E_+^{\prime}+J_+)e^{i\Omega}=
|E|^{\prime}{e^{i \al} \over \sin \bi} e^{i\Omega} \,\, , \,\,\,\,
(E_-^{\prime}+J_-)e^{i\Omega}=
|E|^{\prime}{e^{-i \al} \over \sin \bi} e^{i\Omega}
\ee
Substituting into relation (\ref{interm}) the delta function with respect
to the magnitude of the electric field will force the relation 
$|E(x)|^{\prime}= k\de (x-x_0)$, so that we get the following 
equations
\bq
k \left.{e^{i\al}\over\sin\bi}\right|_{x_0}\!\W+\left.J_+\right|_{x_0} \!\W&=&
-{\sigma_+ \over 2} \W
\label{j+}
\\
k \left.{e^{-i \al} \over \sin \bi }\right|_{x_0} 
\!\W + \left.J_-\right|_{x_0} \!\W&=&
-{\sigma_- \over 2} \W
\label{j-}
\\
\left.J_3\right|_{x_0} \!\W &=&-{\sigma_3 \over 2} \W
\label{j3}
\eq
Relation (\ref{j3}) gives the $\al(x_0)$ dependence of the $\W$ 
components, while in order (\ref{j+}) and (\ref{j-}) to be compatible
we need to choose $k=\pm1/2$.
The unique solutions for the system are
\[
\W=
\left(
\begin{array}{ccc}
\left.e^{-i\al/2} \cos {\bi \over 2}\right|_{x_0}\\ \\
\left.e^{i\al/2} \sin {\bi \over 2}\right|_{x_0}
\end{array}
\right)
\,\, \text{for}\,\, k=-1/2\,\, \text{and}\,\,\,\,
\W=
\left(
\begin{array}{ccc}
\left.-e^{-i\al/2} \sin {\bi \over 2}\right|_{x_0}\\ \\
\left.e^{i\al/2} \cos {\bi \over 2}\right|_{x_0}
\end{array}
\right)
\,\, \text{for}\,\, k=1/2 .
\]
Also equations (\ref{j+}) or (\ref{j-}) could tell us that the sign 
in front of the matrix $\sigma$ in the Gauss' law has to be negative.

\begin{figure}
\centerline{
\put(5,130){$|E|$}
\put(270,15){$x$}
\put(-4,105){$\rho+k$}
\put(18,65){$\rho$}
\put(85,15){$x_0$}
\put(183,15){$x_1$}
\psannotate{\psboxscaled{450}{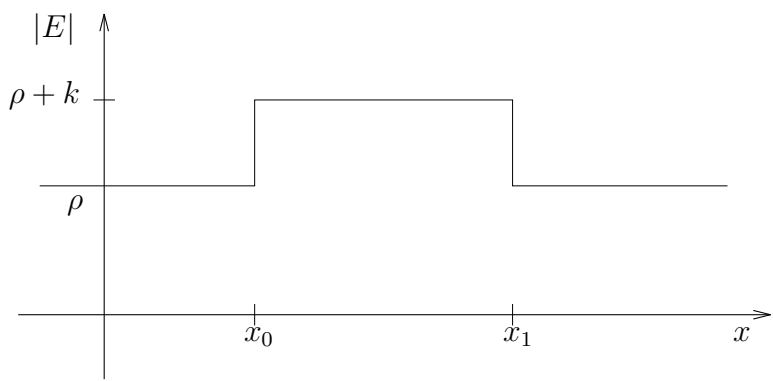}}
{\at(-6\pscm;-1\pscm)
{\caption{\label{ele}The magnitude of the Electric field, $|E|$.}}}
}
\end{figure}

In the case we want to solve the Gauss' law with two conjugate sources
\be
\hat \G_a\Psi[E]=-T_a \Psi[E] \de (x-x_0)+
\Psi[E]T_a \de (x-x_1)
\ee
we need to insert in the wave functional the following delta function
\be
\prod _x\de \Big(|E|-k(\th (x-x_0) -\th (x-x_1))-\rho\Big)
\ee
which ``forces'' the magnitude of
the electric field to take the form as in Figure 
(\ref{ele}).
As now the part of the
wave functional with $x_1$ dependence multiplies the 
source-matrix from the
left, it needs to be a $(1 \times 2)$ row and can be found to be the
hermitian conjugate of the $\W$ solutions found above.
Following the same 
steps as before we see that the complete wave functional for, 
e.g. $k=- 1/2$, is given by
\bq
&&
\Psi[E]=\prod_x \de \Big(|E(x)|+{1 \over 2}
(\th(x-x_0)-\th (x-x_1))-\rho\Big)\times 
e ^{i \int dx |E|\al ^{\prime} \cos\bi} \times 
\no \no
&&
\Sp \Sp
\left(\begin{array}{cc}
 \cos {\bi(x_0) \over 2}\cos {\bi(x_1) \over 2}
e^{-{i \over 2} (\al(x_0)-\al(x_1))}&
\,\, \cos {\bi(x_0) \over 2}\sin {\bi(x_1) \over 2}
e^{-{i \over 2} (\al(x_0)+\al(x_1))}
 \\ \\
 \sin {\bi(x_0) \over 2}\cos {\bi(x_1) \over 2}
e^{{i \over 2} (\al(x_0)+\al(x_1))}&
\,\, \sin {\bi(x_0) \over 2}\sin {\bi(x_1) \over 2}
e^{{i \over 2} (\al(x_0)-\al(x_1))}
\end{array}
\right)
\eq

\section{The Static Quarks Potential}

\begin{figure}
\centerline{
\put(-10,90){$|E|^2$}
\put(-5,60){$k^2$}
\put(85,10){$x_0$}
\put(205,10){$x_1$}
\put(250,10){$x$}
\psannotate{\psboxscaled{550}{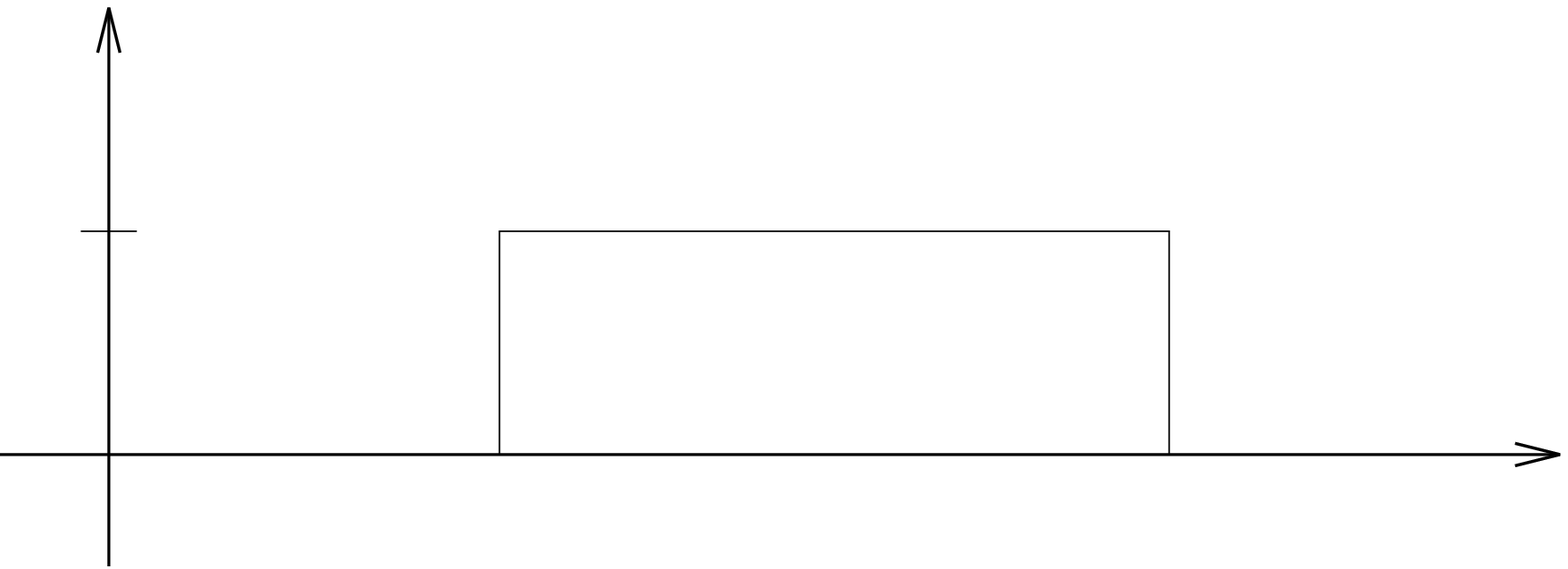}}
{\at(-6\pscm;-1\pscm)
{\caption{
\label{ele12}$|E|^2$
after minimization of the energy ($\rho=0$).}}}
}
\end{figure}

The potential energy of the configuration of the two static sources
is given by
\be
{\cal E} ={1 \over 2} \int dx \int \D E \,\,\tr 
\left( \Psi^{+}[E] \Psi[E] \right) E_a^2(x)-
{1 \over 2}
\int dx \int \D E \,\, \tr \left( \Psi^{+}_v[E] \Psi_v[E] \right) E_a^2(x)
\ee
where $\Psi_v[E]$ is the vacuum functional (without sources)  
\be
\Psi_v[E]=\prod_x \de(|E(x)|-\rho)\times e^{i\Omega} \, .
\ee
Using the relations $\W^+\! \cdot \W=1$ and $(e^{i\Omega})^+e^{i\Omega}=1$ 
we get
\be
{\cal E} =
{3 \over 2} \,g^2\! \int dx \big(k^2 +2k\rho \big)
(\th (x-x_0)-\th(x-x_1))
\ee
where we have reintroduced $g$ in order to restore the dimensionality
of the energy.
The factor $3$ is due to the summation over all the directions of the 
matrix $K$.
If $k=1/2$ then in order to minimize the energy we need 
to choose $\rho=0$ which makes the vacuum energy zero and
\be
{\cal E} = {3 \over 8}\, g^2 (x_1-x_0)
\ee
which is an attractive potential proportional to the distance of the sources
(see Figure (\ref{ele12}a)).
For $k=-1/2$ we should choose $\rho=1/2$ so that
\be
{\cal E}= -{3 \over 8}\, g^2 (x_1-x_0)
\ee
which is a repulsive potential (see Figure (\ref{ele12}b)), 
but the vacuum energy becomes infinite. These potentials can be reproduced in the ``A''-representation by special configurations of Wilson lines; the first one with a path joining the two sources, while the second, with two semi-infinite paths connecting $x_0$ with $- \infty$ and $x_1$ with $+ \infty$ (see \cite{kosy}). The attractive potential gives the well known confinement result. As we have seen, the repulsive one needs an infinite subtraction to give a finite answer, due to the non-vanishing field configuration at infinity, so it does not have a physical analog.

\section{Sources in a General Representation}

We can work out the case of sources belonging to a general representation 
of $SU(2)$ quite easily with the use of the Wigner functions. Let us study the 
Gauss' law of the form
\be
\left( E_a^{\prime}-i\vep_{abc} E_b {\de \over \de E_c} \right) \Psi[E]=
- T_a \Psi[E] \de (x-x_0)
\label{g1}
\ee
where $T_a$'s belong to the $j$ representation of $SU(2)$.
The generalization to many sources is easy as the sum of many sources
at different points in space will lead to a direct product of the 
corresponding columns for each point.
For $E$ belonging to any representation of $SU(2)$ we can decompose it in
general as
$E=gKg^{-1}$ for $g$ an element of $SU(2)$ group and $K$ a matrix belonging 
to the algebra. For a solution of the form 
\be
\Psi[E]=\prod_x \de \Big(K(x)-\tilde K \th (x-x_0)\Big) \times e^{i\Omega} 
\times \W(E(x_0))
\ee
with the real phase $\Omega={i \over c} \int \tr (E g^{\prime} g^{-1})$,
relation (\ref{g1}) becomes
\bq
&&
\prod_x\de \Big(K(x)-\tilde K \th (x-x_0)\Big)\times \W \times e^{i\Omega} \times
\left(E_a^{\prime}+\vep_{abc}E_b{\de \Omega \over \de E_c}\right)- \Sp \Sp
\no
&&
\Sp \Sp\prod_x\de \Big(K(x)-\tilde K \th (x-x_0)\Big)\times e^{i\Omega} \times
i\vep_{abc}E_b{\p \W \over \p E_c}\de (x-x_0)=
\no \no
&&
\Sp=- T_a \prod_x
\de \Big(K(x)-\tilde K \th (x-x_0)\Big)\times e^{i\Omega}\W \de (x-x_0)
\label{g2}
\eq
The term in the parenthesis on the left hand side gives
\be
E_a^{\prime}(x)+\vep_{abc}E_b{\de \Omega \over \de E_c(x)}=
{ 1 \over c}\, \tr(gK^{\prime}g^{-1}T_a)-
{i \over c}\,\tr \!\left( K^{\prime}g^{-1} 
\vep_{abc}E_b{\p g \over \p E_c (x)}\right)
\ee
Substituting into (\ref{g2}) we finally have the following equation
evaluated at the point $x_0$
\be
{1 \over c} \,\W \,\,\tr \! \left(\tilde K g^{-1}\left( T_a g +
J_a g \right)\right) 
+\left( T_a \W + J_a\W\right)=0 \,\, .
\label{g3}
\ee
For solving this equation with respect to $\W$ we shall use the 
properties of the Wigner functions, 
$\D_{m^{\prime}m}^j(\al,\bi,\ga)$ (see e.g. \cite{ang}). 
They represent a general element of the $SU(2)$ group in the general $j$ 
representation. In addition they satisfy the following properties. For the
differential generators, ${\cal J}_a$, of the $SU(2)$ group
\bq
{\cal J}_+&=&e^{i\al}\left( i \cot \bi {\p \over \p \al}
+{\p \over \p \bi } -{i \over \sin \bi}{\p \over \p \ga} \right)
\no
{\cal J}_-&=&e^{-i\al}\left( i \cot \bi {\p \over \p \al}
-{\p \over \p \bi } -{i \over \sin \bi}{\p \over \p \ga} \right)
\no
{\cal J}_3&=&-i {\p \over \p \al}
\eq
which give the angular momentum of a rigid body, we have
\be
{\cal J}_a \D^j(\al,\bi,\ga)=-T_a \D^j(\al,\bi,\ga)
\label{here}
\ee
as well as
\bq
&&
{\cal J}^2 \D _{m^{\prime} m}^j (\al,\bi,\ga)=
j(j+1)\D _{m^{\prime} m}^j (\al,\bi,\ga)\,\, ,
\no \no
&&
{\cal J}_3\D _{m^{\prime} m}^j (\al,\bi,\ga)=-
m^{\prime}\D _{m^{\prime} m}^j (\al,\bi,\ga)\,\, ,
\no \no
&&
{\cal P} _3\D _{m^{\prime} m}^j (\al,\bi,\ga)=-
m \D _{m^{\prime} m}^j (\al,\bi,\ga)
\eq
for ${\cal J}^2=({\cal J}_+{\cal J}_-+{\cal J}_-{\cal J}_+)/2+{\cal J}_3^2
={\cal J}_1^2+{\cal J}_2^2+{\cal J}_3^2$ and 
${\cal P}_3=-i \p / \p \ga$.
For the particular representation of $T _a$ satisfying (\ref{here}),
we have that $c=j(j+1)(2j+1)/3$.
The first step will be to identify the general element, $g$, of the $SU(2)$
with the Wigner function $\D _{m^{\prime} m}^j (\al,\bi,\ga)$.
As in relation (\ref{g3}) the operators $J_a$ do not include differentiation
with respect to $\ga$, it is convenient to write the operators ${\cal J}_a$ as
\be
{\cal J}_a =J_a +\Gamma_a {\cal P}_3
\ee
for $\Gamma_a$ being zero for $a=3$ and $e^{\pm i\al}/\sin \bi$ for $a=\pm$,
so that the following equation holds
\be
\left.(T_a) \right._{m^{\prime}n}\D _{n m}^j (\al,\bi,\ga)+
J_a \D _{m^{\prime} m}^j (\al,\bi,\ga)=
\Gamma_a m \D _{m^{\prime} m}^j (\al,\bi,\ga)
\ee
and equation (\ref{g3}) becomes
\bq
&&
{1 \over c}\,\Gamma_a\de _{m^{\prime}m} \tilde K_{m^{\prime}n} 
\left.\D^{-1}\right. _{n k}^j (\al,\bi,\ga)\,
m \D _{k m}^j (\al,\bi,\ga) \W+
\left( T_a \W + J_a\W\right)=0\Rightarrow
\no \no
&&
\Sp \Sp \Sp {1 \over c}\,\Gamma_a \!\sum_{m=-j}^{j} m\tilde K _{mm}\W+
\left(T_a  \W + J_a\W\right)=0
\eq
The matrix $\tilde K$ could be any constant element of the algebra, so 
via a gauge transformation we can make it equal to
$kT_3$ and we eventually have
$\sum_{m=-j}^{j} m\tilde K _{mm}=$ $k \, j(j+1)(2j+1)/3$ and
\be
k \, \Gamma_a \W+
\left( T_a  \W + J_a\W\right)=0
\ee
which is solved for a column vector $\W^{(s)}$ of the form
\be
\W^{(s)}=\D _{m s}^j (\al,\bi,\ga\!=\!0)
\label{www}
\ee
for a given fixed $s$.
This fixes also the value of $k$ to be $k=-s$.
For example in the case of the fundamental representation 
we have $j=1/2$ and $s=\pm 1/2$ so that $k=\mp1/2$, 
which agrees with the previous result.

We can calculate for the wave functional $\Psi ^{(s)}[E]=\exp (i \Omega) 
\prod_x \de \Big(|E(x)|+k
(\th(x-x_0)-\th (x-x_1))-\rho\Big) \W^{(s)}(x_0)\otimes \W^{(s)+} (x_1)$, the string tension, 
$\sigma$ for the static 
potential, ${\cal E}$, in the general representation $j$, i.e.
\be
{\cal E} =\sigma^{(s)} (x_1-x_0) \,\, .
\ee
As the $\D$ matrices are unitary,
$\W^{(s)+} \! \cdot \W^{(s)}=1$, 
so the same consideration as in the previous section 
is applied. If $k$ is greater than zero,
then $\rho$ should be zero and the string tension is $\sigma^{(s)}=
{3 \over 2} \,g^2 s^2$
or in the case $k$ is less than zero then $\rho=-k$ and
$\sigma^{(s)}=-{3 \over 2}\, g^2 s^2$.
We see that $\sigma^{(s)}$, as well as the particular solution $\W^{(s)}$, 
depends on the representation in which 
the sources belong and are characterized by $s$ ranging from $-j$ to $j$.

It is interesting to see how these solutions are related with the 
well familiar Wilson line. In the ``A''-representation 
we know that the path ordered exponential
\be
W[A]\equiv P e^{i \,g\!\int ^{x_1}_{x_0} dx A_a T_a }
\label{wili}
\ee
satisfies the Gauss' law with two conjugate sources in the 
$j$ representation. Using the Hamiltonian 
(\ref{arep}) we find that the effective potential
of the system of the two quarks is 
\be
{\cal E}_A={1 \over 2} \,g^2 \,C_2 \, (x_1-x_0)
\ee
where $C_2$ is the quadratic Casimir operator given by
$C_2=j(j+1)$ in the $j$ representation.
It is easy to find which functional in the ``E''-representation gives the same string tension as the Wilson line. We notice the similarity of our setup to the quantum mechanical angular momentum, where $E$ is interpreted as the angular momentum vector $J$, and the static potential ${\cal E}$ (apart from the $x$ dependence) as the expectation value of $J^2$ with respect to the state $\mid \! s, J\rangle$, where $s$ is the magnetic quantum number. In order to obtain the expectation value $j(j+1)$ from $J^2$ we need to take the average state over all spatial directions of $J$, which is equivalent with an average over all values of $s$, i.e. $\sum _s \mid \! s,J\rangle$. 
In this spirit we define in the ``E''-representation the average $\bar\Psi$
as
\be
\bar \Psi[E] \equiv \sum_{s =-j} ^j \Psi^{(s)} [E]
\ee
where the index $s$ labels the independent solutions with different 
polarizations. Using the orthogonality of the states $\Psi^{(s)}$,
it is easy to derive the effective potential for the $\bar \Psi$ state
\be
{\cal E}_E ={1 \over 2}\, g^2 \,C_2 \, (x_1-x_0) \,\, .
\ee
The string tension is also proportional to the quadratic Casimir operator
which is an important common feature of the two representations
\cite{Green}. In addition we see that in the case we take $x_1\equiv x_0$
the solutions in the two representations are both 
proportional to the identity.
Therefore we conjecture that $\bar \Psi$ is the functional Fourier 
transformation of the Wilson line (\ref{wili}). Finally, we point out that in the procedure of solving the Gauss' law in the ``E''-representation with sources, we were forced to choose for $\W$ a column decomposition of $\D$ and not the whole $\D$ matrix as happens in the ``A''-representation, which gave us $2j+1$ independent solutions.

\section{Acknowledgments}
J. P. would like to thank Roman Jackiw and Kenneth Johnson
for helpful conversations.

\vspace{0.1in}

\vspace{0.3in}

\begin{figure}
\centerline{
\put(-10,90){$|E|^2$}
\put(-5,60){$k^2$}
\put(85,10){$x_0$}
\put(205,10){$x_1$}
\put(250,10){$x$}
\psannotate{\psboxscaled{550}{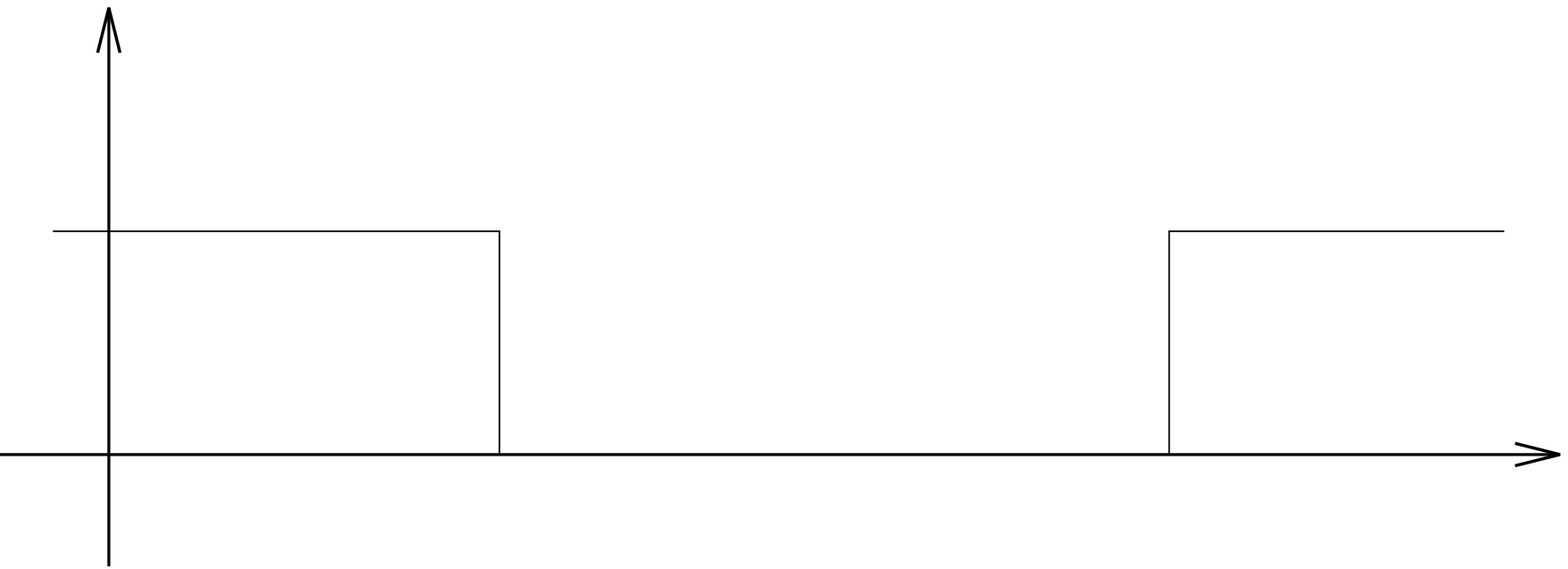}}
{\at(-6\pscm;-1\pscm)
{\caption{
\label{ele21}$|E|^2$
after minimization of the energy ($\rho=-k$).}}}
}
\end{figure}

\end{document}